**Spin injection into Si in three-terminal vertical and four-terminal lateral devices with Fe/Mg/MgO/Si tunnel junctions having an ultrathin Mg insertion layer**

Shoichi Sato[1], Ryosho Nakane[1,2], Takato Hada[1], and Masaaki Tanaka[1,3]

[1]*Department of Electrical Engineering and Information Systems, The University of Tokyo, 7-3-1 Hongo, Bunkyo-ku, Tokyo 113-8656, Japan*
[2]*Institute for Innovation in International Engineering Education, The University of Tokyo, 7-3-1 Hongo, Bunkyo-ku, Tokyo 113-8656, Japan*
[3]*Center for Spintronics Research Network (CSRN), The University of Tokyo, 7-3-1 Hongo, Bunkyo-ku, Tokyo 113-8656, Japan*

We demonstrated that the spin injection/extraction efficiency is enhanced by an ultrathin Mg insertion layer ($\leq$ 2 nm) in Fe/Mg/MgO/$n^+$-Si tunnel junctions. In diode-type vertical three-terminal devices fabricated on a Si substrate, we observed the narrower three-terminal Hanle (N-3TH) signals indicating true spin injection into Si, and estimated the spin polarization in Si to be 16% when the thickness of the Mg insertion layer is 1 nm, whereas no N-3TH signal was observed without Mg insertion. This means that the spin injection/extraction efficiency is enhanced by suppressing the formation of a magnetically-dead layer at the Fe/MgO interface. We have also observed clear spin transport signals, such as non-local Hanle signals and spin-valve signals, in a lateral four-terminal device with the same Fe/Mg/MgO/$n^+$-Si tunnel junctions fabricated on a Si-on-insulator substrate. It was found that both the intensity and linewidth of the spin signals are affected by the geometrical effects (device geometry and size). We have derived analytical functions taking into account the device structures, including channel thickness and electrode size, and estimated important parameters; spin lifetime and spin polarizations. Our analytical functions well explain the experimental results. Our study shows the importance of suppressing a magnetically-dead layer, and provides a unified understanding of spin injection/detection signals in different device geometries.

# I. INTRODUCTION

Si-based spin-transistors, which have ferromagnetic source/drain and a Si channel, have generated much attention, since they are very attractive for building-blocks in next-generation integrated circuits [1 – 3]. Spin transistors can be used for nonvolatile memory and reconfigurable logic circuits, because their transistor characteristics can be changed by the magnetization configuration of the ferromagnetic source and drain. To realize their functions, we need large magnetoresistance in the source-channel-drain transport, which requires 1) efficient spin injection/extraction of spin-polarized electrons into/from a Si channel (so-called "spin injection/extraction"), and 2) efficient transport of spin-polarized electrons via the Si channel (so-called "spin-dependent transport"). Recently, spin MOSFET operation at room temperature was reported, but the magnetoresistance ratio $\gamma_{MR}$ was very small (0.02 – 0.12%) [4, 5]. If the spin injection/extraction efficiency is greatly enhanced and spin-dependent transport via a Si channel becomes coherent, much larger $\gamma_{MR}$ will be obtained in spin MOSFET operation. Although spin injection/extraction and spin-dependent transport in Si channels have been studied so far [5 – 11], the physics and detailed mechanism remain unclear. Moreover, it has been theoretically pointed out that the device geometry can affect the spin injection/extraction signals (hereafter "geometrical effect") [12 – 16], but it is not experimentally verified yet. Recently, we have analyzed broader three-terminal Hanle (B-3TH) signals (which are *not* originated from *true* spin injection [17 – 21]) obtained by the three-terminal method [15, 22], and have proposed a model [21] (hereafter "dead layer model") suggesting that the coherency of electron spins is reduced by a magnetically-dead layer formed at a ferromagnetic metal/oxide interface. Note that the "dead layer" can be an ultrathin (one atomic layer or less) paramagnetic layer or paramagnetic interface states. Our "dead layer model" predicts that the spin injection/extraction efficiency will be enhanced if we can eliminate such a dead layer.

In this study, we show that the spin injection efficiency is enhanced in Fe/Mg/MgO/Si junctions by inserting an ultrathin Mg layer (thickness $t_{Mg} \leq 2$ nm) between the ferromagnetic Fe layer and the MgO tunnel barrier. This enhancement is attributed to the suppression of the magnetically-dead layer at the Fe/MgO interface, which is verified by the shape of B-3TH signals and magnetization measurements. It is noteworthy that the Mg insertion between Fe and MgO is a well-known technique in magnetic tunnel junctions (MTJs) to suppress the formation of a dead layer and to improve the spin coherence of tunneling electrons [23]. In order to analyze spin injection/extraction signals correctly, we prepared two types of device structures with different geometries. One device structure is named "vertical device" (shown in Fig. 1), which is a tunnel diode structure with a circular electrode patterned on a bulk Si substrate. This structure allows us to estimate accurate spin polarization and spin lifetime by the narrower three-terminal Hanle (N-3TH) signals (which are originated from *true* spin injection [17, 22]), because it is not necessary to take into account the geometrical effect. The other device structure is named "lateral device" (shown in Fig. 5(a)), which has a thin-body Si channel with four electrodes patterned on a Si-on-insulator (SOI) substrate. This structure allows us to prove the true spin injection in the Si channel by the four-terminal (non-local) measurements [24]. However, the geometrical effect must be taken into account in the analysis.

In section II, we investigate how the Mg insertion affect spin injection/extraction in the vertical devices and un-processed junction structures with various Mg-layer thicknesses ($t_{Mg} = 0 - 2$ nm). N-3TH and B-3TH signals observed in the vertical devices are changed depending on $t_{Mg}$. By analyzing the B-3TH signals using the dead layer model, we found that the dead layer at the Fe/MgO interface is suppressed by the Mg insertion, leading to the true spin injection/extraction into/from Si. The suppression of the dead layer is supported by the magnetization measurements of the un-processed samples. A relatively high spin polarization $P$ = 16% in Si is obtained when the Mg-layer thickness is 1 nm.

In section III, we verified the realization of spin transport and pure spin current in a Si channel using a lateral device with an 1 nm-thick Mg insertion layer. We observed four-terminal Hanle (4TH) signals, four-terminal spin-valve signals, and N-3TH signals in the lateral device. To analyze the experimental results, we derived analytical functions which take into account the effect of the channel thickness and the electrode lengths. By comparing the spin injection/extraction signals in the both vertical and lateral devices, we experimentally show that the geometrical effect must be taken into account for the precise analysis of spin injection/detection and spin transport.

## II. Mg INSERTION IN THE VERTICAL DEVICES

A. Sample preparation

Figure 1 shows a schematic illustration of the vertical device with (from top to bottom) an Al(~160 nm)/Mg(1 nm)/Fe(3 nm)/Mg($t_{Mg}$)/MgO($t_{MgO}$)/$n^+$-Si(001) junction and an Al backside contact, where the Mg insertion layer thickness $t_{Mg}$ is 0 − 2 nm and the MgO tunnel barrier thickness $t_{MgO}$ is 0.5 − 1.2 nm. The fabrication process is as follows: First, a phosphorus-doped $n^+$-Si ($8\times10^{19}$ cm$^{-3}$) substrate with H-terminated surface was thermally cleaned at 900°C for 15 min in an ultra-high vacuum chamber (base pressure ~ $3\times10^{-7}$ Pa). Then, an MgO layer was deposited on the surface by electron beam evaporation at 30°C with a rate of 0.003 nm/sec. It is noteworthy that the MgO layer was not crystallized from reflective high-energy electron diffraction (RHEED) observation. Subsequently, without breaking vacuum the substrate was transferred into a molecular beam epitaxy chamber (MBE) via a vacuum transfer chamber, then Mg/Fe/Mg multilayers and an Al (~ 10 nm) cap layer were successively deposited at room temperature by using Knudsen cells. Here, the bottom Mg layer ($t_{Mg}$ nm) was inserted to prevent the reaction of the Fe layer and the MgO tunnel barrier and thereby to suppress the formation of a dead layer, and the top Mg layer (1 nm) was inserted to prevent the reaction of the Fe layer and the Al cap layer [25]. Then, immediately after being exposed to air, a 160-nm-thick Al layer was deposited on the surface and top electrodes with diameter $d$ = 5.6 and 17.8 μm were fabricated by UV lithography and $H_3PO_4$ etching for many junctions. Finally, an Al layer was deposited on the backside of the substrate just after removing native oxide by Ar ion milling and HF etching. The junction area for *I-V* measurements was 25 μm$^2$ ($d$ = 5.6 μm), whereas that for spin injection/extraction measurements by the three-terminal method was 250 μm$^2$. ($d$ = 17.8 μm).

B. Magnetization of the non-processed sample

To measure the magnetic properties, a non-processed sample having the same layered structure with $t_{Mg}$ = 0 – 2 nm and $t_{MgO}$ = 2 nm was also prepared as a reference. Figure 2(a) shows $t_{Mg}$ dependence of the saturated magnetization ($M_S$) which was estimated from magnetization *vs*. in-plane magnetic field (*M – H*) curves measured at 4 K by a superconductive quantum interference device (SQUID) magnetometer. Since $M_S$ increases with increasing $t_{Mg}$ and saturates at $t_{Mg}$ ≥ 1 nm, the formation of a magnetically-dead layer at the Fe/MgO interface was suppressed by the Mg insertion layer with $t_{Mg}$ ≥ 1 nm. Considering that $M_S$ = 1230 emu/cc at $t_{Mg}$ = 0 nm and $M_S$ = 1450 emu/cc at $t_{Mg}$ = 2 nm, the thickness of the dead layer at the Fe/MgO interface was estimated to be 0.3 nm when $t_{Mg}$ = 0 nm. Since the dead layer is probably FeO$_x$ [26], the constant $M_S$ in $t_{Mg} \geq 1$ nm indicates that the Fe layer does not touch with the bottom MgO layer, namely, the Mg layer fully covered the bottom MgO layer.

C. Experimental results of the *I-V* characteristics and three-terminal measurements

All the samples show non-linear *I-V* curves as in Fig. 2(b) ($t_{MgO}$ = 0.8 nm, $t_{Mg}$ = 1 nm) and the resistance-area product (*RA*) at zero bias show a exponential dependence on $t_{MgO}$, as shown in Fig. 2(c) ($t_{Mg}$ = 0 and 2 nm). This indicates that tunnel current via the MgO barrier layer is dominant in our devices. From Simmons' equation [27] and Fig. 2(c), the barrier height of MgO ($\Phi_{MgO}$) was estimated for each $t_{Mg}$ and plotted in Fig. 2(d); as $t_{Mg}$ increases from 0 to 2 nm, $\Phi_{MgO}$ (= 0.29 eV) decreases at first and then becomes almost constant (0.11 eV) between 1 and 2 nm. Since the work functions of Fe and Mg are 4.7 and 3.7 eV [28], respectively, it is most likely that the barrier height is decreased by inserting the Mg layer between Fe and MgO. Combining the data of Fig. 2(a) and Fig. 2(d), the decrease in $\Phi_{MgO}$ results from the increase in the coverage of Mg over the MgO layer at $t_{Mg} < 1$ nm, and the constant $\Phi_{MgO}$ at $t_{Mg} \geq 1$ nm results from the full coverage of Mg over the MgO layer. In consequence, the increase in $M_S$ is correlated with the decrease in $\Phi_{MgO}$, and $t_{Mg}$ = 1 nm is the lowest thickness for obtaining the high $M_S$ and the low $\Phi_{MgO}$ at the same time.

Figure 1 shows our three-terminal measurement setup, in which the junction voltage drop $V_{3T}$ was measured by a voltmeter, while a constant current *I* was driven from the top electrode to the backside of the substrate and an external magnetic field *H* was applied sweeping from −3000 Oe to 3000 Oe along the in-plane ($\theta$ = 0°) or normal-to-plane ($\theta$ = 90°) directions. Note that distance between the injection/extraction electrode and reference electrode is at least 1300 μm, which is much longer than expected spin diffusion length (~ 1 μm) [10]. Figures 3(a) – (d) show the change in three-terminal signals $\Delta V_{3T}(H,\theta)$ of the samples with $t_{MgO}$ = 0.8 nm and $t_{Mg}$ = 0, 0.5, 1, and 2 nm, respectively, which were measured at 4 K with $I$ = −30 mA (the spin extraction regime). In Fig. 3(a) – (d), the red and green curves correspond to $\theta$ = 0° and $\theta$ = 90° conditions, respectively. Broader three-terminal Hanle (B-3TH) signals ($\theta$ = 90°) [17, 21] and inverted three-terminal Hanle (I-3TH) signals ($\theta$ = 0°) [29] were observed at $H$ = −3000 Oe ~ 3000 Oe in all the sample, and their amplitudes decreased as $t_{Mg}$ increased. Note that the N-3TH signals (*true* spin injection signal) [17, 22] were observed at $H$ = −300 Oe ~ 300 Oe at $\theta$ = 90° in the samples with $t_{Mg} \geq$ 0.5 nm, as shown in the inset of Figs. 3(b) – (d), whereas no N-3TH signal was observed in the sample with $t_{Mg}$ = 0 nm.

It should be also noted that although N-3TH signals were observed in the spin *extraction* regime with $I$ = −30 mA as shown in the insets of Figs. 3(b) – (d), no clear N-3TH signal was observed but only B-3TH signals were observed in the spin *injection* regime with $I$ = +30 mA (not shown here), as reported previously [30, 31]. This difference in the N-3TH result due to the $I$ polarity can be explained by the electric field in the Si channel at the MgO/Si interface [32]. In the spin extraction regime ($I$ < 0), the electric field in the Si channel is almost screened by the accumulated electrons caused by the high *n*-type doping concentration $8 \times 10^{19}$ cm$^{-3}$ in Si [33], namely, the electrical potential in Si is almost flat. Thus, the electric field effect is negligible in the spin extraction regime, and the electron spins in Si are purely diffusive. On the other hand, in the spin injection regime ($I$ > 0), a depletion layer is formed in Si nearby the MgO/Si interface and the electron spins injected into Si are drifted away from the interface by the electric field in the depletion layer. Using one-dimensional (1-D) Poisson's equation, the maximum electric field strength in the Si depletion layer in our sample was estimated to be 3.4 MV/cm when $I$ = +30 mA, and this electric field decreases the amplitude of the N-3TH signal down to ~ 14% and broadens the linewidth by ~ 3000%, compared with those in the spin extraction regime (see Supplemental Materials (S.M.) [34]). Thus, it is reasonable that no clear N-3TH signal appears in the spin injection regime, probably because such weak and broadened N-3TH signals cannot be distinguished from the intense B-3TH signals even if it exists. So far, disappearance of spin injection signals in three-terminal devices with a Si channel was reported [30, 31], but the reason has not been clarified. This is probably because the spin injection/extraction signals were analyzed using the simple 1-D spin diffusion equation which was established in all metallic systems, *i.e.*, semiconducting properties of the Si channel have not been taken into account in the analysis so far. As we suggested here, the electric field and the depletion layer differ in the Si channel between the injection and extraction conditions, even if the doping concentration of Si is significantly high. Thus, this electric field effect is important to design semiconductor-based spintronic devices using spin injection and detection.

D. Analysis of the three-terminal signals

Considering that the three-terminal signals $\Delta V_{3T}(H,\theta)$ are the superposition of the N-3TH signal $\Delta V^{N-3TH}(H,\theta)$ and the B-3TH signal $\Delta V^{B-3TH}(H,\theta)$, we analyzed the signals in Figs. 3(a) – (d) with the equation $\Delta V_{3T}(H,\theta) = \Delta V^{N-3TH}(H,\theta) + \Delta V^{B-3TH}(H,\theta)$, as we carried out the same procedure in our previous study [21]. Since the electrode diameter ($d$ = 17.8 μm) is much larger than the expected spin diffusion length ($\lambda_S$ ~ 1 μm) [10] and the electric field in the Si is negligible in the spin extraction regime as mentioned earlier, the 1-D spin diffusion model is applicable. The B-3TH and N-3TH functions for the vertical device ($\Delta V^{N-3TH(vertical)}$) are as follows [17, 21, 22, 35] (see S.M. for details [34]):

$$\Delta V^{B-3TH}(H,\theta) = \eta_{B-3TH} V_0 \frac{(H\cos\theta + S)^2 + C^2}{(H\cos\theta + S)^2 + (H\sin\theta)^2 + B^2 + C^2}, \tag{1}$$

$$\Delta V^{N-3TH(vertical)}(H,\theta) = \Delta V_0^{spin}\left[\sqrt{\frac{1+\sqrt{1+(\gamma\tau_S H)^2}}{2+2(\gamma\tau_S H)^2}}\sin^2\theta + \cos^2\theta\right], \tag{2}$$

$$\Delta V_0^{spin} = J\rho\lambda_S P^I P^D, \tag{3}$$

where $J$ is the current density, $P^I$ is the spin polarization of elections injected into Si (hereafter "injection polarization"), $P^D$ is the spin polarization detected by the detection electrode (hereafter "detection polarization"), $H$ is the external applied field, and $\theta$ is the field angle, $\gamma$ is the gyromagnetic ratio, $\tau_S$ is the spin lifetime in Si, $\lambda_S$ is the spin diffusion length, $\rho$ is the Si resistivity, $V_0$ is the offset voltage drop of the tunnel junction at $H = 0$, and $\eta_{B-3TH}$ is the B-3TH ratio [21]. Ideally, when spin injection/extraction efficiency is 100%, $P^I$ and $P^D$ are the same as the spin polarization of Fe electrode (~ 40%) [36]. Parameters $C$, $B$, and $S$ in Eq. (1) are the effective internal magnetic fields in the ultrathin magnetically-dead layer introduced in our previous study (see S.M. of ref. [21]). $S$ is the directional field parallel ($S > 0$) or antiparallel ($S < 0$) to the magnetization $M_{Fe}$ of the Fe layer, $C$ is the non-directional field parallel to $M_{Fe}$, and $B$ is the non-directional field perpendicular to $M_{Fe}$. Parameter $B$ is the primary indicator of a magnetically-dead layer and is strongly related to the I-3TH signal. As $B$ decreases, the amplitude of I-3TH signal decreases. When the paramagnetic state completely vanishes and ferromagnetic order appears in a magnetically-dead layer, the I-3TH signal disappears and $B = 0$. In the analysis, we use $\gamma = 1.76\times10^7$ s$^{-1}$Oe$^{-1}$ and assumed that the spin injection and detection polarizations are the same value $P_{3T}$, that is, $P_{3T} = P^I = P^D$. It is notable that Eq. (2) is twice as large as the conventional N-3TH functions [9]. This is because injected spins diffuse vertically down to the backside of the substrate in this device structure, whereas they diffuse laterally both to the left and the right side in the lateral devices on a SOI substrate (see S.M. [34]).

First, the B-3TH signals were analyzed since these must be subtracted from $\Delta V_{3T}(H,\theta)$ to extract and analyze the N-3TH signals. Black solid curved in Figs. 3 (a) – (d) are fitting results using Eq. (1). By fitting Eq. (1) to the experimental results in Figs. 3(a) – (d), $B$, $C$, $S$, and $\eta_{B-3TH}$ were estimated and plotted in Fig. 4(a) and (b). It was found that $S$ and $C$ increase and $B$ decreases as $t_{Mg}$ increases, and this means that the magnetically-dead layer thickness is decreased and ferromagnetic order increased at the Fe/Mg/MgO interface with increasing $t_{Mg}$ [21]. Considering this result with the fact that $S$ is positive except at $t_{Mg} = 0$ nm, the ferromagnetic order appeared and the formation of the dead layer was suppressed at $t_{Mg} \geq 0.5$ nm. This is consistent with the result that the N-3TH signals were only observed at $t_{Mg} \geq 0.5$ nm (see the insets of Figs. 3(b) – (d)). Also, as shown in Fig. 4(b), $\eta_{B-3TH}$, which is the ratio of the amplitude of B-3TH and I-3TH signals to the tunnel voltage drop, also decreased with increasing $t_{Mg}$. Thus, $B$, $C$, $S$, and $\eta_{B-3TH}$ are correlated with each other and they are also correlated with $M_S$ in Fig. 2(a), as expected.

Then, the N-3TH signals in the insets of Figs. 3(b) – (d) were analyzed by fitting Eqs. (2) and (3) to the experimental signals with the measurement parameters $J = 0.12$ A/μm$^2$, $\rho = 1.0$ mΩcm, and $\lambda_S = 1$ μm taken from ref. [10]. Black solid curved in the insets of Figs. 3 (b) – (d) are fitting results. Figures 4(c) and (d) show the estimated $\tau_S$ and $P_{3T}$, respectively. The spin lifetime $\tau_S \sim 2$ ns obtained in Fig. 4(c) for all $t_{Mg}$ is reasonable because $\tau_S$ is determined only by the decaying precession of electron spins in the Si substrate and is independent of the junction properties; the spin polarization of electrons at the Si surface is not related to $\tau_S$. Moreover, this $\tau_S$ value ~ 2 ns is consistent with the previously reported values (1 – 10 ns) [8, 10, 37, 38] and in good agreement with the theoretically calculated value (2.5 ns) for the same phosphorus concentration in Si [39]. On the other hand, $P_{3T}$ = 8 – 16% in Fig. 4(d) is comparable with the previously reported values 5 – 17% [10 – 11]. Contrary to $\tau_S$, $P_{3T}$ changes depending on $t_{Mg}$; $P_{3T}$ ~16% for $t_{Mg}$ = 0.5 and 1.0 nm, and $P_{3T}$ ~ 8% for $t_{Mg}$ = 1.5 and 2.0 nm. Although the dead layer formation was significantly suppressed when $t_{Mg} \geq$ 1.0 nm, the injected electron spins lost their polarization while passing through the Mg layer when $t_{Mg} \geq$ 1.5 nm. Thus, we concluded that $t_{Mg}$ = 1.0 nm is the best condition for our spin injection/detection junctions.

### III. SPIN DEPENDENT TRANSORT IN THE LATERAL DEVICE

A. Sample preparation

To confirm the spin injection into the Si layer and the spin transport in the Si channel and also to explore the geometrical effect [12 – 16] on spin-related signals, a lateral device structure was fabricated on an SOI substrate, as shown in Figs. 5(a) and (b), and compared the spin transport properties with those of the vertical devices of Fig. 1. Figures 5(a) and (b) are schematic (a) side-view and (b) top-view illustrations of the lateral device having the same junction structure as that in the vertical device of Fig. 1(a) with $t_{Mg}$ = 1.0 nm and $t_{MgO}$ = 0.8 nm. The fabrication process is as follows: First, an undoped SOI substrate was doped with phosphorus by the thermal diffusion method with a $P_2O_5$ film on the surface. After removing the $P_2O_5$ film and successive cleaning with $H_2SO_4$ solution, H-terminated surface was formed with HF. Then, tunnel junctions were formed by the same procedure as that for the vertical device. After being exposed to air, a 100-nm-thick Ta layer was deposited on the surface and electrodes were formed by EB lithography and Ar ion milling. Finally, each device was isolated by etching the Si body layer with $CF_4$ gas, as shown Figs. 5(a) and (b). The channel length $L_{ch}$ and width $W_{ch}$ are 2 and 180 μm, respectively, the lengths $l_A$ and $l_B$ along the $y$ direction of the electrode A and electrode B (the inside two electrodes) were 1 and 5 μm, respectively. The outside electrodes R1 and R2 with $l_R$ = 40 μm in length along the $y$ direction are the reference electrodes, and the distance $L_{ref}$ between the electrodes A and R1 (B and R2) was ~ 100 μm. Since the Si channel resistivity was ~ 1 mΩcm from the $I$-$V$ characteristics, the electron carrier density was estimated to be ~ $1\times10^{20}$ cm$^{-3}$. Thus, $L_{ref}$ = 100 μm is much longer than the expected spin diffusion length of ~ 1 μm in Si with this doping concentration [10]. Figures 5(c) and (d) show our four-terminal measurement setup I and II, respectively. Here, in setup I, we define the four-terminal voltage $V_{4T}$ and three-terminal voltage $V_{3T}$ as follows; $V_{4T}$ is the voltage between electrodes B and R2, and $V_{3T}$ is the voltage between A and R2. We measured $V_{3T}$ and $V_{4T}$ while a constant current ($I$ = ±50 mA) was driven from electrode A to R1 and an external magnetic field was applied sweeping between −3000 Oe and +3000 Oe along the in-plane (the $x$ axis, $\theta$ = 0°) or normal-to-plane (the $z$ axis, $\theta$ = 90°) directions. $V_{4T}$ and $V_{3T}$ in the setup II are also defined by exchanging the connection A with B, and R1 with R2.

B. Experimental results of four-terminal measurements

Figure 6(a) shows the change $\Delta V_{4T}$ in $V_{4T}$ measured in setup I with an in-plane magnetic field ($\theta$ = 0°) and $I$ = −50 mA (the spin extraction regime), where the red and blue curves represent the major and minor loops, respectively. Since the major loop of $\Delta V_{4T}$ shows two minimum plateaus between 50 – 80 Oe and −50 – −80 Oe, which reasonably agree with the coercivities of the electrodes A and B, this is the spin-valve signal. Figures 6(b) and (c) show $\Delta V_{4T}$ signals in the parallel and antiparallel magnetization configurations measured in setup I with a normal-to-plane magnetic field ($\theta$ = 90°) and $I$ = −50 mA, where the red and blue curves represent the signals measured in the parallel and antiparallel magnetization configurations, respectively. Almost the same signals as in Figs. 6(a–c) with inverted polarity are observed with $I$ = 50 mA (spin injection regime, not shown). Clear Hanle signals with the change in polarity depending on the magnetization configuration give strong evidence that spin-polarized electrons are transported via the Si channel [24]. Moreover, the amplitude of the signal change ~ 2 μV in Fig. 6(a) is nearly equal to the sum of the signal changes observed in Figs. 6(b) and (c), indicating that the data of Fig. 6(a) are caused by the spin-valve effect.

On the other hand, the red curves in Figs. 7(a) and the blue curves in Fig. 7(b) are $\Delta V_{3T}$ at $\theta$ = 0° measured in setup I and setup II, respectively. Besides, the red and blue curves in Fig. 7(c) are the same signals between ±300 Oe in Figs. 7(a) and (b), respectively. The hysteresis characteristics of these signals probably come from the tunnel anisotropic magnetoresistance (TAMR) [40], since the signals in setup I and II show the minimum values at the coercivities of electrode A (±80 Oe) and electrode B (±50 Oe), respectively. We considered that the $\Delta V_{3T}$ signal at $\theta$ = 0° is composed of both this hysteretic TAMR signal and the I-3TH signal, and fitted Eq. (1) to estimate the I-3TH signal, as illustrated by the black curve in Figs. 7(a) and (b). As in the case of the vertical device, the amplitude of the I-3TH signal was small compared with the B-3TH signal, which confirms the suppression of the dead layer at the Fe/Mg/MgO interface of electrode A and B, by inserting an ultrathin Mg layer between the Fe and MgO layers.

We also show $\Delta V_{3T}$ signals measured with a normal-to-plane magnetic field ($\theta$ = 90°) in setup I and II by the green and brown curves in Fig. 7(a) and (b), respectively, where the fitting results of the B-3TH signals using Eq. (1) are also shown by the black solid curves (details will be described later). Comparing the $\Delta V_{3T}$ signals in the vertical (Fig. 3(c)) and lateral (Figs. 7(a) and (b)) devices with the same $t_{Mg}$ and $t_{MgO}$, the amplitude of the N-3TH and the TAMR signals in the lateral device is several times larger than that in the vertical devices (details will be discussed later), while the amplitude of the B-3TH and I-3TH signals in both devices are almost the same. The same amplitude of the B-3-TH and I-3TH signals in the two types of device structures confirms again that the B-3TH and I-3TH signals do not originate from spin accumulation in the Si channel but from the magnetoresistance depending on the tunnel junction properties.[17 – 21, 38]. If we use the same function Eq. (2), which was derived from the 1-D spin diffusion model, the larger N-3TH signals observed in the lateral device lead to inconsistent fitting results $P_{3T}$ = 20% and $\tau_S$ = 1.0 ns (electrode A), and $P_{3T}$ = 63% and $\tau_S$ = 2.3 ns (electrode B), although $P_{3T}$ and $\tau_S$ should be comparable in both electrodes and also to those in the vertical device with $t_{Mg}$ = 1 nm ($P_{3T}$ = 16% and $\tau_S$ = 1.7 nsec). This means that spin accumulation signals in the lateral devices must be analyzed by a more sophisticated model with taking into account the geometrical effects, when the geometrical scale of the structure, such as the SOI channel thickness and electrode length, is smaller than $\lambda_S$. The geometrical effects on the spin accumulation signals were pointed out by other groups [12 – 16], but have never been experimentally verified. In this study, we use two device structures with different geometries, and thus we can clarify what determines the shape and amplitude of the N-3TH signal by comparing the N-3TH signals in these two device structures. On the contrary, the difference of the TAMR signals in the vertical and lateral devices probably reflects the magnetization switching process of the Fe electrodes in each device because the shape of the Fe electrodes in each device is different. It is reported that the TAMR signal is proportional to both the tunnel resistance and the vertical component of the magnetization vector [40]. Considering that the tunnel area resistance ($RA$) is about 10 kΩμm$^2$ and the amplitude of the TAMR in the lateral device (Figs. 7(a) and (b)) is 0.5 – 2 Ωμm$^2$, the difference of the TAMR signals in the vertical and lateral devices can occur when the vertical component of the Fe magnetization changed by ~ 0.02% during the magnetization switching process. Such a change of the Fe magnetization switching process is possible because it is strongly affected by the shape of the ferromagnets.

C. Analysis of the four-terminal signals

Based on the two-dimensional (2-D) spin diffusion model and the ideas in refs. [12 – 14], we originally constructed the following analytic functions for $\Delta V_{3T}$ and $\Delta V_{4T}$ in the lateral device ($\Delta V^{\text{N-3TH(lateral)}}$ and $\Delta V^{\text{4TH(lateral)}}$), with taking into account the injector electrode length $l^I$, detector electrode length $l^D$, and the SOI thickness $t_{\text{SOI}}$ (see S.M. for detailed derivation [34]):

$$\Delta V^{\text{4TH(lateral)}}(H,L) = \Delta V_0^{spin} \frac{1}{2} \frac{\lambda_S}{t_{\text{SOI}}} \text{Re}\left[\frac{1}{1+i\gamma H\tau_S} \exp(-\alpha L_{ch}) \frac{1}{\alpha l^D}\left(1-\exp(-\alpha l^D)\right)\left(1-\exp(-\alpha l^I)\right)\right] \quad (4)$$

for $\Delta V_{4T}$, and

$$\Delta V^{\text{3TH(lateral)}}(H) = \Delta V_0^{spin} \frac{\lambda_S}{t_{\text{SOI}}} \text{Re}\left[\frac{1}{1+i\gamma H\tau_S}\left\{1-\frac{1}{\alpha l^I}(1-\exp(-\alpha l^I))\right\}\right] \quad (5)$$

for $\Delta V_{3T}$,

where $\alpha = \frac{\sqrt{1+i\gamma H\tau_S}}{\lambda_S}$, $i$ is the imaginary unit, and Re[ ] is the real part of the square brackets. In deriving Eq. (4) and (5), we assume that the spin injection (current density) is uniform over the electrode, $t_{\text{SOI}} \ll \lambda_S$, and $W_{ch} \gg \lambda_S$. Here, the factor $\lambda_S/t_{\text{SOI}}$ in Eqs. (4) and (5) is an indicator of the channel confinement effect (CCE), which means that the spin accumulation is significantly larger than that in the vertical device as $t_{\text{SOI}}$ becomes smaller than $\lambda_s$ [12]. Also, the factor $\frac{1}{\alpha l^D}\left(1-\exp(-\alpha l^D)\right)\left(1-\exp(-\alpha l^I)\right)$ in Eq. (4) and $\frac{1}{\alpha l^I}\left(1-\exp(-\alpha l^I)\right)$ in Eq. (5) are indicators of the electrode averaging effect (EAE), which means that averaging the spin detection signals over the detector along the $y$ direction. As $l^I$ and $l^D$ become longer, both amplitudes and linewidths of N-3TH and 4TH signals are changed as follows:

· Amplitude

Amplitudes of 4TH signals become smaller because the average distance between the injection and detection electrodes becomes longer. On the contrary, amplitudes of N-3TH signals become larger because CCE is more pronounced.

· Linewidth

Linewidths of both the N-3TH and 4TH signals become narrower. This means that the injected spins are dephased by a smaller magnetic field because the phase variation of the detected spins becomes larger.

In setup I, $l^I = l_A$ and $l^D = l_B$ are used, and in setup II, $l^I = l_B$ and $l^D = l_A$ are used. From the fitting, $\tau_S$, $\lambda_S$, and the average spin polarization $P_{4T} = \sqrt{P^I P^D}$ are estimated from Eq. (3) and (4), and $\tau_S$, $P_{3T}$ of electrode A ($P_{3T(A)}$), and electrode B ($P_{3T(B)}$) are estimated from Eq. (3) and (5).

In Figs. 6(b) and (c), the fitting results using Eq. (4) and parabolic backgrounds are shown by the black and broken curves, respectively, from which $P_{4T}$ = 7.2%, $\tau_S$ = 2.0 ns, and $\lambda_S$ = 1.0 μm were estimated. From the analysis of $\Delta V_{3T}$, the B-3TH and I-3TH signals were analyzed by Eq. (1), and then the N-3TH signals were analyzed by Eq. (5). The black curves in Figs. 7(a) and (b) show the fitting results for the B-3TH and I-3TH signals, in which $S \sim 350$ Oe, $B \sim 450$ Oe, and $C \sim 1300$ Oe estimated by using Eq. (1) are in good agreement with those estimated in the vertical device with $t_{Mg}$ = 1 nm (see Fig. 4(a)). To see the effect of EAE on the shape of N-3TH signals, the N-3TH signals obtained in the electrode A (setup I) and electrode B (setup II) are shown in Fig. 7(d), where the green and brown curves represent the signals obtained in electrode A and B, respectively. The linewidths of the N-3TH signals in Fig. 7(d) are quite different between both cases, but fitting with Eq. (5) (black curves) leads to comparable values; $\tau_S$ = 1.3 ns for electrode A and 1.7 ns for electrode B. Furthermore, using $\lambda_S$ = 1.0 μm estimated from the 4TH signal (Figs. 6(b)), $P_{3T(A)}$ = 6.6% and $P_{3T(B)}$ = 12% were estimated from the N-3TH signals (Fig. 7(d)). In consequence, the parameters estimated from the N-3TH signals ($\sqrt{P_{3T(A)}P_{3T(B)}}$ = 9.1% and $\tau_S$ = 1.3 − 1.7 ns) are comparable with those from the 4TH signals ($P_{4T}$ = 7.2% and $\tau_S$ = 2.0 ns). This result confirms again that both the N-3TH and 4TH signals come from the true spin accumulation in Si. Moreover, since these values estimated from the 4TH (Fig. 6(b)) and N-3TH signals (Fig. 7(d)) in the lateral device are close to those estimated from the N-3TH signals (Fig. 3(c)) in the vertical device ($P_{3T}$ = 16% and $\tau_S$ =1.7 ns), it is quite reasonable to conclude that Eqs. (4) and (5) precisely express the spin accumulation signals under the geometrical effects; CCE and EAE. Therefore, these equations are appropriate for accurate estimation of $P_{3T}$, $P_{4T}$, and $\tau_S$ in lateral device structures.

D. Comparison of the fitting results with/without the geometrical effects

To confirm this conclusion, we fitted the following three sets of equations (i) − (iii) and parameters ($P_{3T}$, $P_{4T}$, $\tau_S$) listed in Table 1 to the N-3TH signals observed in both the vertical and lateral devices (Figs. 3(c) and 7(d)) and the 4TH signals observed in the both setup I (Fig. 6(b)) and II (not shown): (i) Eqs. (4) and (5) (both CCE and EAE are taken into account), (ii) Eqs. (4) and (5) with $l^I, l^D \to 0$ (without EAE, Eqs. (S20) and (S21) in S.M. [34]), and (iii) Eq. (4) with $\lambda_S/t_{SOI} = 1$ and $l^I, l^D \to 0$ and Eq.(2) (without CCE and EAE, Eq. (3) in ref. [10]). The estimated values ($P_{3T}$ = 16% and $\tau_S$ = 1.7 ns) from the N-3TH signals in the vertical device (shown in the second and third columns of Table 1) were identical, because both the electrode length (17.8 μm) and channel thickness (675 μm) is much larger than $\lambda_S$ = 1.0 μm. Also, $P_{4T}$ and $\tau_S$ estimated from the 4TH signals with $I$ = +50 mA (the spin injection regime) in the both setup I and II are listed in Table 2. The parameters related to the N-3TH signals are not listed in Table 2 because they were not clearly observed in the spin injection regime ($I > 0$) probably due to the depletion layer formation as mentioned before. Note that the 4TH signal was not observed in setup II with $I$ = −50 mA (marked by ND in Table 1), although the 4TH signal was observed with both bias polarities in setup I. This probably comes from the unwanted electric field concentration [41] at the left edge of the electrode B (the side closer to the electrode R2) and effective channel length becomes longer than $L_{ch}$ so that spin polarized electrons cannot reach the detector electrode (electrode A).

From the fitting results in table 1 and 2, the following features (a – d) are clarified: (a) For the spin polarization, $P$ = 11 – 63% is estimated by (iii), while $P$ = 2.5 – 14% is estimated by (ii). $P$ values are overestimated without CCE, especially when the injector electrode length is longer (setup II). (b) For the spin lifetime, $\tau_S$ = 2.1 – 3.2% is estimated by (ii), while $\tau_S$ = 1.7 – 2.3% is estimated by (i). Without EAE, $\tau_S$ values are overestimated especially when the injector or detector length is longer. (c) $P$ = 2.5 – 7.2% is estimated by (ii), while $P$ = 7.2 – 12% is estimated by (i) in 4TH. Without EAE, $P$ values in 4TH are underestimated especially when the detector length is longer (setup I). (d) With CCE and EAE, variation of the estimated values reduced from ($\tau_S$ = 1.0 – 3.2 ns and $P$ = 11 – 63%) to ($\tau_S$ = 1.3 – 2.3 ns and $P$ = 6.6 – 12%), and these values become close to those in the vertical device ($\tau_S$ = 1.7 ns and $P$ = 16%). From these features (a) – (d), we concluded that both CCE and EAE must be taken into account for the precise analysis of the N-3TH and 4TH signals in the thin channel device structure.

## IV. CONCLUSION

First, we investigated magneto-transport properties of Fe/Mg/MgO/Si tunnel junctions (vertical device) with various Mg insertion layer thicknesses ($t_{Mg}$) by three-terminal Hanle measurements. The formation of a magnetically-dead layer at the Fe/MgO interface was prevented by inserting a ultrathin Mg layer ($t_{Mg} \geq 0.5$ nm) between Fe and MgO. The highest spin polarization $P = 16\%$ was achieved when $t_{Mg} = 1$ nm. These results are consistent with our previously proposed model, which suggests that a magnetically-dead (paramagnetic) layer formed at the ferromagnetic metal / oxide interface causes B-3TH and I-3TH signals and reduces the spin injection polarization. This is the first study that experimentally shows the relationship between true spin injection/extraction signals (N-3TH) and other B-3TH and I-3TH signals.

Then, realization of spin injection/extraction and pure spin current was verified by the observation of both the four-terminal spin-valve effect and the four-terminal Hanle effect using the lateral device structure with $t_{Mg} = 1$ nm. The fitting functions were originally derived from the 2-D spin diffusion model, taking into account the geometrical effects, CCE and EAE. Using the fitting functions with the geometrical effects, the $\tau_S$ and $P$ values were estimated in both the vertical and lateral devices and they are in good agreement, whereas these values estimated using the functions without the geometrical effects were not in agreement between the vertical devices and lateral devices. These results indicate that the geometrical effects must be taken into account for the precise estimation of the spin lifetime $\tau_S$ and spin polarization $P$.

To realize spin transistors with highly spin-dependent output characteristics, further understanding and control of the spin injection/extraction efficiency are needed. This work provides a universal procedure to analyze the spin injection/detection signals observed both in vertical and lateral devices and will contribute to the precise understanding of the physics concerning spin injection/extraction and spin transport in semiconductor device structures.

ACKNOWLEDGEMENTS

This work was partially supported by Grants-in-Aid for Scientic Research, including Specially Promoted Research, Project for Developing Innovation Systems of MEXT, Yazaki Science and Technology Foundation, and Spintronics Research Network of Japan. Part of this work was carried out under the Cooperative Research Project Program of RIEC, Tohoku University.

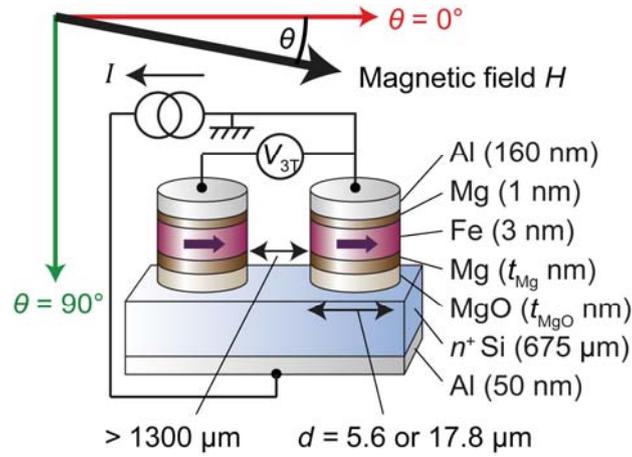

Figure 1  Schematic illustration of the vertical device having a Fe(3 nm)/Mg($t_\text{Mg}$ nm)/MgO($t_\text{MgO}$ nm)/$n^+$-Si tunnel junctions.  Three-terminal measurement setup is also shown.  Constant current $I$ is driven from the top to the backside and three-terminal voltage ($V_\text{3T}$) is measured while an external magnetic field is applied (−3000 Oe ~ 3000 Oe).  The magnetic field direction $\theta$ is varied from 0° to 90°; $\theta = 0°$ and $\theta = 90°$ are the in-plane and normal-to-plane directions, respectively.  Distance between the injection/extraction electrode and reference electrode is at least 1300 μm.

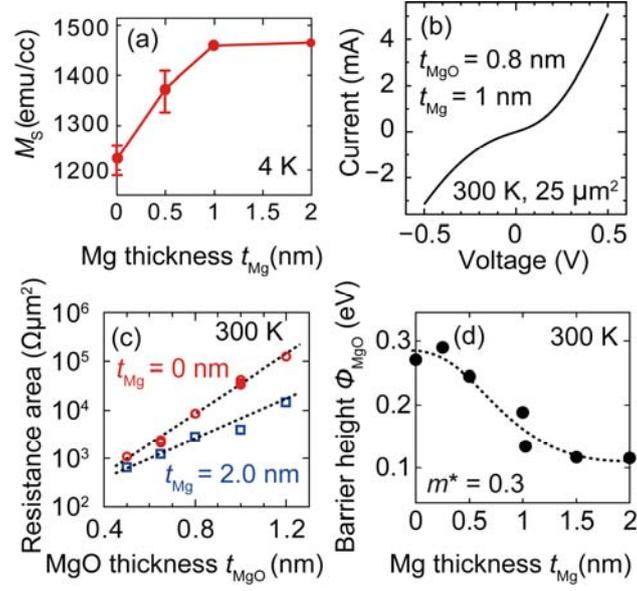

Figure 2 (a) Saturation magnetization $M_S$ of non-processed samples having the same layered structure as in Fig. 1 with various $t_{Mg}$ (= 0, 0.5, 1.0, and 2.0 nm) and $t_{MgO}$ = 2 nm, which were measured at 4 K with an in-plane magnetic field of 20 – 30 kOe. (b) *I-V* characteristic measured at 300 K of the vertical device with $t_{Mg}$ = 1 nm and $t_{MgO}$ = 0.8 nm. (c) Resistance area (*RA*) at *V* = 0 estimated from the *I-V* characteristics plotted as a function of $t_{MgO}$. From the dotted lines, the MgO barrier height $\Phi_{MgO}$ was estimated to be 0.27 eV for $t_{Mg}$ = 0 nm and 0.11 eV for $t_{Mg}$ = 2.0 nm. (d) $\Phi_{MgO}$ plotted as a function of $t_{Mg}$, in which the broken curve is a guide for eyes.

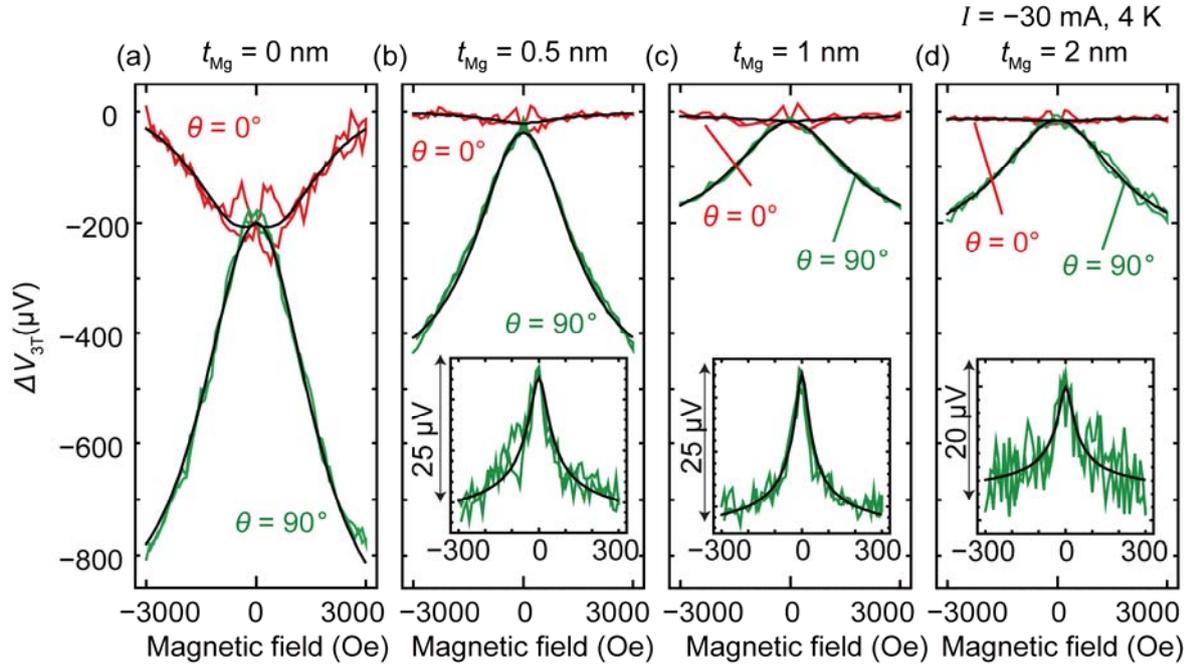

Figure 3 (a) – (d) Change of the three-terminal Hanle signals $\Delta V_{3T}$ measured at 4 K with $I = -30$ mA and in-plane magnetic field ($\theta = 0°$) and normal-to-plane magnetic field ($\theta = 90°$) for the sample with various Mg thicknesses; (a) $t_{Mg} = 0$ nm, (b) $t_{Mg} = 0.5$, (c) $t_{Mg} = 1.0$, and (d) $t_{Mg} = 2.0$ nm. Red and green curves represent the signals for the in-plane ($\theta = 0°$) and the normal-to-plane ($\theta = 90°$) magnetic field, respectively, and black solid curves are the fitting results using Eq. (1). Insets of (b) – (d) show the N-3TH signals in a lower field range (−300 Oe ~ 300 Oe) after subtracting the B-3TH signals.

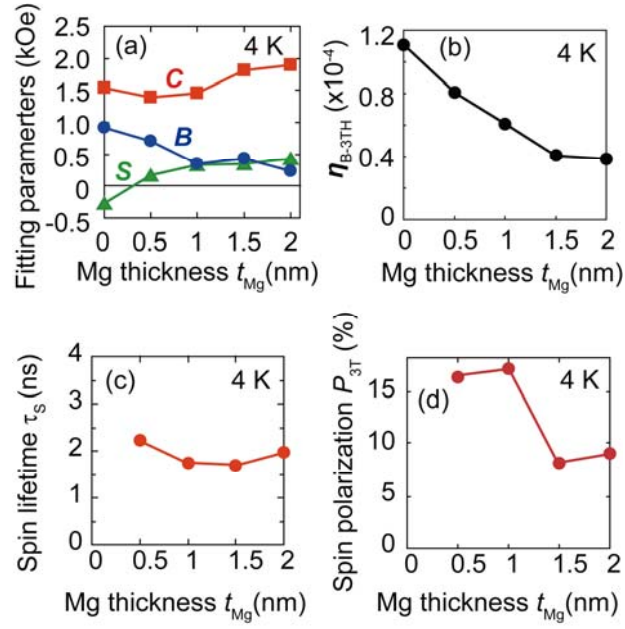

Figure 4 (a)(b) Mg thickness ($t_{Mg}$) dependence of the fitting parameters (a) $B$, $C$, $S$ and (b) $\eta_{B\text{-}3TH}$ estimated by using Eq. (1) and the experimental B-3TH and I-3TH signals in Figs. 3(a) – (d). Blue circles, red squares, and green triangles in (a) represent the values for $B$, $C$, and $S$, respectively. (c)(d) $t_{Mg}$ dependence of the fitting parameters $\tau_S$ and $P_{3T}$ (= $P^I = P^D$ was assumed) estimated by using Eqs. (2) and (3) and the N-3TH signals in the insets of Figs. 3(b) – (d).

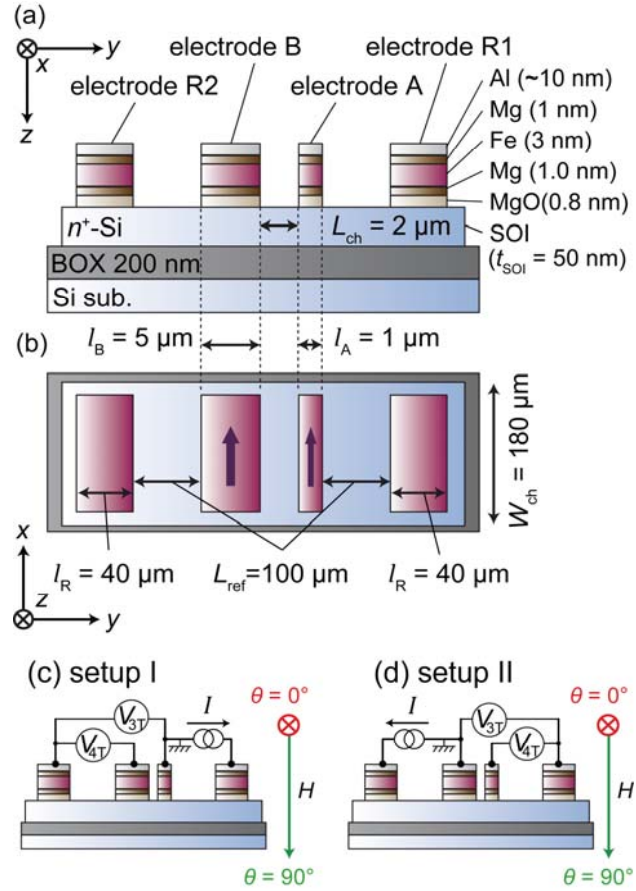

Figure 5 (a) Side view and (b) top view of the vertical device having Fe(3 nm)/Mg(1 nm)/MgO(0.8 nm)/$n^+$-Si tunnel junctions fabricated on a silicon-on-insulator (SOI) substrate for four-terminal measurements. Four electrodes are named R2, B, A, and R1 from the left to right. Coordinates are defined as follows; $x$ and $y$ are parallel to the long and short sides of the electrodes, respectively, and $z$ is normal to the substrate plane. (c)(d) Four-terminal measurement (c) setup I and (d) setup II, where the three-terminal signal $V_{3T}$ and four-terminal signal $V_{4T}$ are measured at the same time while an external magnetic field is applied (−3000 Oe ~ 3000 Oe). The magnetic field direction $\theta$ is varied from 0° to 90°; $\theta = 0°$ and $\theta = 90°$ are the in-plane and normal-to-plane directions, respectively.

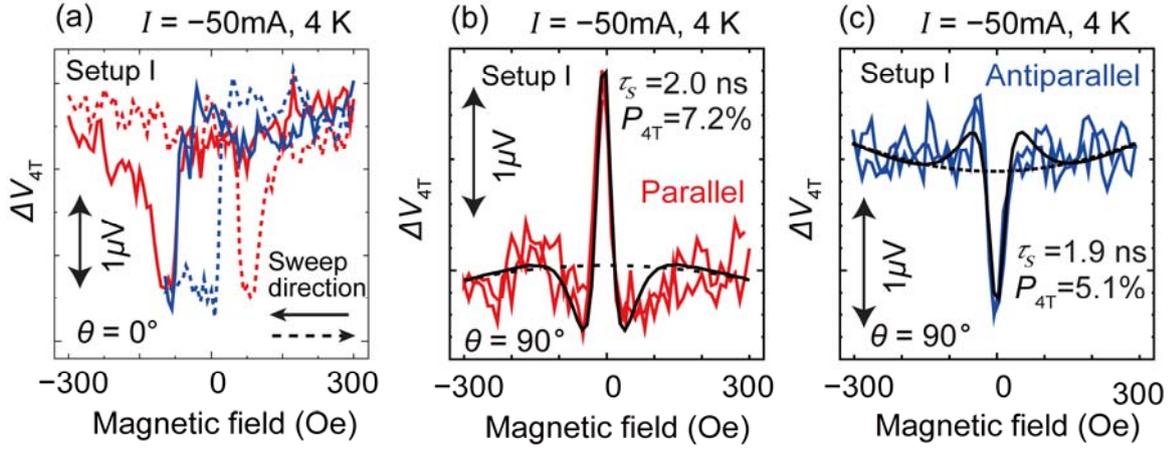

Figure 6 (a) Change in the four-terminal signal $\Delta V_{4T}$ as a function of in-plane magnetic field ($\theta = 0°$) indicating the spin-valve effect, measured at 4 K with $I = -50$ mA in setup I. Red solid/dashed and blue solid/dashed curves are major and minor loops, respectively. (b)(c) Four-terminal Hanle signals $\Delta V_{4T}$ as a function of normal-to-plane magnetic field ($\theta = 90°$) measured at 4 K with $I = -50$ mA in the (b) parallel and (c) antiparallel magnetization configurations. Black solid and black dashed curves are the fitting results with Eq. (4) and parabolic backgrounds, respectively.

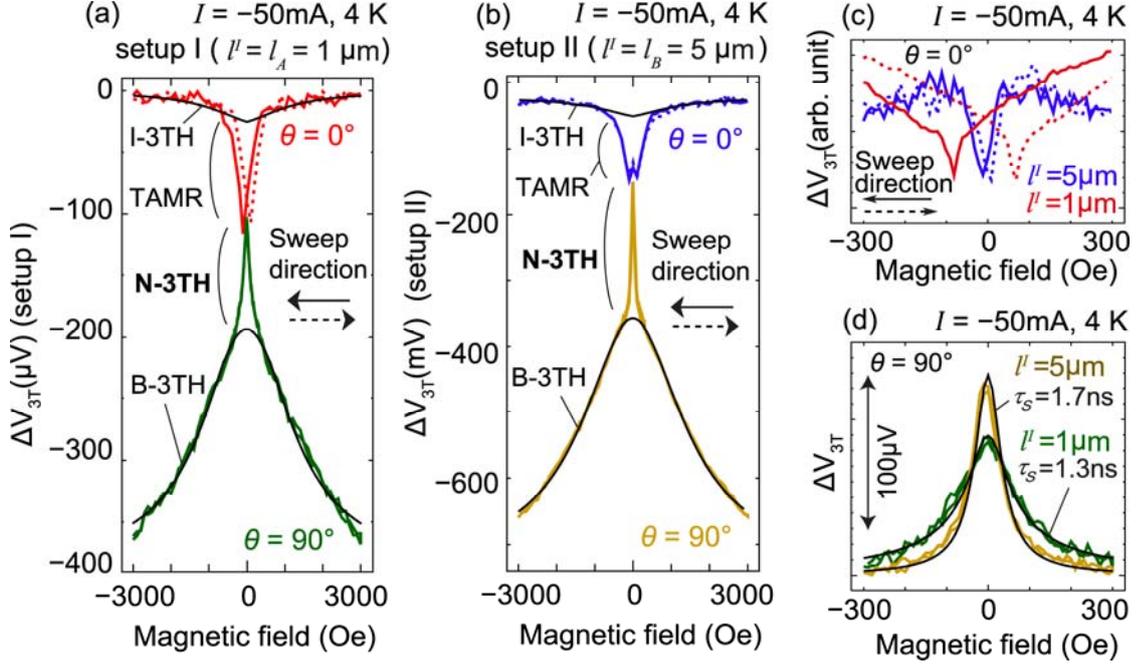

Figure 7 (a)(b) Change in the three-terminal signals $\Delta V_{3T}$ measured at 4 K with $I = -50$ mA in (a) setup I and (b) setup II, where red and blue curves represent the signals for the in-plane ($\theta = 0°$), and green and brown curves represent the signals for the normal-to-plane ($\theta = 90°$) magnetic field, respectively. Fitting curves of the I-3TH and the B-3TH signals with Eq. (1) are also shown by black solid curves. (c) $\Delta V_{3T}$ signals ($\theta = 0°$) within ±300 Oe, where solid/broken red curves and solid/broken blue curves are the signals in (a) and (b), respectively. (d) Normal-to-plane N-3TH signals ($\theta = 90°$) after subtracting B-3TH within ±300 Oe, where green and brown curves are the experimental curves in (a) and (b), respectively. Black solid curves are fitting results with Eq. (5).

Table 1  $P_{3T}$, $P_{4T}$, and $\tau_S$ estimated from the experimental signals with $I = -50$ mA (spin extraction regime) in Figs. 3(c), 6(b), and 7(d), using Eqs. (2) – (5) under various conditions; (i) Eqs. (4) and (5) (both CCE and EAE are taken into account), (ii) Eqs. (4) and (5) with $l_I, l_D \to 0$ (without EAE), and (iii) Eqs. (4) and (5) with $\lambda_S / t_{SOI} = 1$ and $l_I, l_D \to 0$ (without CCE and EAE). ND denotes absence of signals.

| Device | Vertical | | Lateral | | | | | | | |
|---|---|---|---|---|---|---|---|---|---|---|
| Method | N-3TH | | N-3TH (setup I) | | N-3TH (setup II) | | 4TH (setup I) | | 4TH (setup II) | |
| Value | $P_{3T}$ | $\tau_S$ | $P_{3T}$ | $\tau_S$ | $P_{3T}$ | $\tau_S$ | $P_{4T}$ | $\tau_S$ | $P_{4T}$ | $\tau_S$ |
| (i) | 16% | 1.7 ns | 6.6% | 1.3 ns | 12% | 1.7 ns | 7.2% | 2.0 ns | ND | ND |
| (ii) | 16% | 1.7 ns | 4.5% | 1.0 ns | 14% | 2.3 ns | 2.5% | 2.1 ns | ND | ND |
| (iii) | 16% | 1.7 ns | 20% | 1.0 ns | 63% | 2.3 ns | 11% | 2.1 ns | ND | ND |

Table 2  $P_{4T}$ and $\tau_S$ estimated from the experimental signals with $I = +50$ mA (spin injection regime), using Eqs. (5) under the same conditions as in Table 1.

| Device | Lateral | | | |
|---|---|---|---|---|
| Method | 4TH (setup I) | | 4TH (setup II) | |
| Value | $P_{4T}$ | $\tau_S$ | $P_{4T}$ | $\tau_S$ |
| (i) | 7.5% | 1.9 ns | 12% | 2.3 ns |
| (ii) | 2.7% | 2.3 ns | 7.2% | 3.2 ns |
| (iii) | 12% | 2.3 ns | 32% | 3.2 ns |

Supplemental Material

Spin injection into Si in three-terminal vertical and four-terminal lateral devices with Fe/Mg/MgO/Si tunnel junctions having an ultrathin Mg insertion layer

Shoichi Sato[1], Ryosho Nakane[1,2], Takato Hada[1], and Masaaki Tanaka[1,3]

[1]*Department of Electrical Engineering and Information Systems, The University of Tokyo, 7-3-1 Hongo, Bunkyo-ku, Tokyo 113-8656, Japan*

[2]*Institute for Innovation in International Engineering Education, The University of Tokyo, 7-3-1 Hongo, Bunkyo-ku, Tokyo 113-8656, Japan*

[3]*Center for Spintronics Research Network (CSRN), The University of Tokyo, 7-3-1 Hongo, Bunkyo-ku, Tokyo 113-8656, Japan*

## S1. Derivation of the 3-dimensional impulse response for the vertical device

In this section, to take into account the geometrical effects for the spin accumulation functions, first we find the impulse response of the 3-dimentinal (3-D) spin diffusion function (Eq. S1) in the space domain and obtain general solutions by integrating it. Here, we consider electron spins injected into a semi-infinite channel ($z \geq 0$) from an infinitesimal area $dx_0 dy_0$ at $(x,y,z)=(x_0,y_0,0)$ (see Fig. S1). The injected electron spins at $(x,y,z)$ under an external magnetic field $\boldsymbol{H}$ are described by the following differential equation [S1];

$$D\nabla^2 \boldsymbol{S} - \frac{1}{\tau_S}\boldsymbol{S} - \gamma \boldsymbol{H} \times \boldsymbol{S} + 2\boldsymbol{P}^1 \frac{Jdx_0 dy_0}{q}\delta^3(x-x_0, y-y_0, z) = 0, \quad (S1)$$

where $\boldsymbol{S} = (S_x, S_y, S_z)$ is the spin density, $\boldsymbol{P}^1 = (P^1,0,0)$ is the injector polarization, $J$

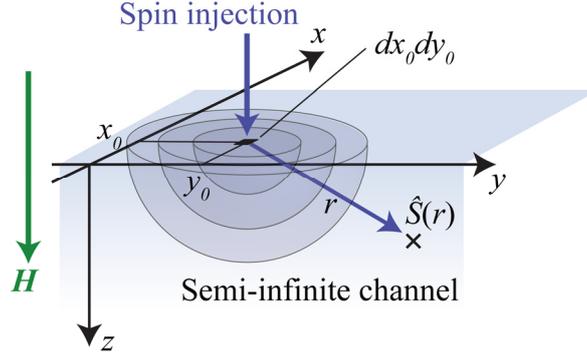

Figure S1　Geometry for our 3-D spin diffusion model in vertical devices.　Electron spins are injected into a semi-infinite channel ($z \geq 0$) from an infinitesimal area $dx_0 dy_0$ at $(x, y, z) = (x_0, y_0, 0)$. $\hat{S}(r)$ represents the complex spin density at a distance $r$ from the spin injecting point ($x_0$, $y_0$, 0).

is the injection current density, $q$ is the elementary charge, $D$ is the diffusion constant of electrons, $\tau_S$ is the spin lifetime, $\gamma$ is the gyromagnetic ratio, and $\delta^3(x, y, z)$ is the 3-D Dirac delta function.　The first, second, third, and fourth terms in the left side of Eq. (S1) express spin diffusion, spin relaxation, spin precession, and spin injection, respectively.　The factor 2 in the fourth term of the left side in Eq. (S1) comes from the boundary condition that injected electron spins do not flow into the region $z < 0$. $S_i$ ($i = x, y, z$) is defined by the difference between the densities of up and down electron spins $S_i = n_\uparrow - n_\downarrow$, where $n_\uparrow$ and $n_\downarrow$ are the electron densities of up and down spins, respectively, when the quantization axis is along the $i$ direction.　When an external magnetic field is applied parallel to the $z$ axis $\boldsymbol{H} = (0, 0, H)$, Eq. (S1) can be written by

$$D\nabla^2 S_x - \frac{1}{\tau_S} S_x + \gamma H S_y + 2 P^I \frac{J dx_0 dy_0}{q} \delta^3(x - x_0, y - y_0, z) = 0, \quad \text{(S2a)}$$

$$D\nabla^2 S_y - \frac{1}{\tau_S} S_y - \gamma H S_x = 0, \quad \text{(S2b)}$$

$$D\nabla^2 S_z - \frac{1}{\tau_S} S_z = 0 \quad \text{(S2c)}$$

From Eq. (S2c), $S_z = 0$ is obtained.　Introducing the complex spin density

$\hat{S} = S_x + iS_y$, Eqs. (S2a) and (S2b) are converged into the following single deferential equation:

$$D\nabla^2 \hat{S} - (\frac{1}{\tau_S} + i\gamma H)\hat{S} + 2P^I \frac{Jdx_0 dy_0}{q} \delta^3(x - x_0, y - y_0, z) = 0. \quad (S3)$$

Since the system in the channel has the spherical symmetry whose center lies at point $(x_0, y_0, 0)$, we use the spherical coordinate and express the spin density as a function of the distance from the center $r = \sqrt{(x-x_0)^2 + (y-y_0)^2 + z^2}$. Thus, Eq. (S3) is transformed into

$$D\left(\frac{d^2}{dr^2} + \frac{2}{r}\frac{d}{dr}\right)\hat{S}(r) - (\frac{1}{\tau_S} + i\gamma H)\hat{S}(r) + 2P^I \frac{Jdx_0 dy_0}{q} \frac{\delta(r)}{4\pi r^2} = 0, \quad (S4)$$

where $\frac{\delta(r)}{4\pi r^2}$ is the 3-D Dirac delta function in the spherical coordinate. Boundary conditions are $S(r \to \infty) = 0$ and the flux continuity at $r = 0$. The latter is expressed as

$$\left[-2\pi r^2 D \frac{d}{dr} \hat{S}(r)\right]_{r \to +0} = P^I \frac{Jdx_0 dy_0}{q}, \quad (S5)$$

where the left side of Eq. (S5) is the total spin flux passing through the hemisphere of radius $r$ in the channel, and the right side is the total spin injection flux from the infinitesimal area $dx_0 dy_0$. The solution of Eq. (S4) is given by

$$\hat{S}(r, H) = \frac{P^I J}{Dq} \frac{dx_0 dy_0}{2\pi} \frac{\exp(-\alpha r)}{r}, \quad (S6)$$

where $\alpha = \frac{\sqrt{1 + i\gamma H \tau_S}}{\lambda_S}$ and $\lambda_S = \sqrt{D\tau_S}$ is the spin diffusion length. Here we assume that the chemical potential difference between up and down electron spins can be written as $\text{Re}[\hat{S}(r,H)]/N(E_F)$ [S2] using the density of states at the Fermi level $N(E_F)$ and Einstein's relation in a degenerated semiconductor $1/\rho = q^2 N(E_F) D$, where $\rho$ is the channel resistivity. The electric potential of the detector electrode with the polarization $\boldsymbol{P}^D = (P^D, 0, 0)$ at the position $r$ is given by

$$\Delta V(r,H) = P^D \frac{\text{Re}\left[\hat{S}(r,H)\right]}{qN(E_F)} = P^I P^D J\rho\lambda_S \, \text{Re}\left[\frac{dx_0 dy_0}{2\pi\lambda_S} \frac{\exp(-\alpha r)}{r}\right]. \quad (S7)$$

Here, we define the maximum spin signal amplitude $\Delta V_0^{spin}$ and the impulse response function $f^{3D}(r,H)$ for 3-D spin transport as follows;

$$\Delta V_0^{spin} = P^I P^D J\rho\lambda_S, \quad (S8)$$

$$f^{3D}(r,H)dx_0 dy_0 = \frac{dx_0 dy_0}{2\pi\lambda_S} \frac{\exp(-\alpha r)}{r}. \quad (S9)$$

Eq. (S8) is the same as Eq. (3) in the main manuscript. Using Eqs. (S8) and (S9), Eq. (S7) is expressed as

$$\Delta V(r,H) = \Delta V_0^{spin} \, \text{Re}\left[f^{3D}(r,H)dx_0 dy_0\right]. \quad (S10)$$

Eq. (S10) expresses the spin accumulation signal voltage observed at the distance $r$ from the infinitesimal spin injection at $(x_0, y_0, 0)$. Spin signals in any vertical device is calculated by integrating the impulse response $f^{3D}(r,H)$ over both the injector and detector electrode.

### S2. Derivation of the 2-dimentional impulse response for the vertical device

In this section, we derive the two-dimensional (2-D) impulse response for a vertical device having rectangular electrodes with the longitudinal axis along the $x$-axis, as shown in Fig. S2(a). In our lateral device (shown in Fig. 5 in the main manuscript), spin distribution in the channel is approximately uniform along the $x$-axis, so the 2-D impulse response is applicable to our analysis. We consider the situation that electron spins are injected into a semi-infinite channel ($z \geq 0$) from an injector electrode having infinite length along the $x$-axis and finite length $l^I$ along the $y$ axis connecting to the channel at $z = 0$. The injected electron spins diffuse in the channel, and they are detected by the detector electrode having an infinite length along the $x$ axis and a finite length $l^D$ along the y axis connecting to the channel at $z = 0$. The distance between the injector and detector electrodes is $L$.

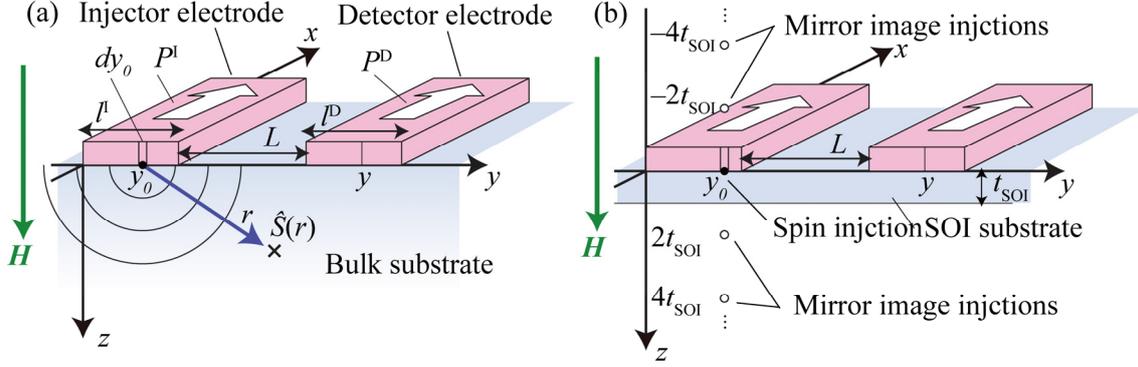

Figure S2 (a)(b) Geometries for our 2-D spin diffusion model in (a) vertical and (b) lateral devices. Both injector and detector electrodes are infinitely long along the $x$ axis, and their lengths along the $y$ axis are $l^I$ and $l^D$, respectively. Spin polarization axes of both electrodes are parallel to the $x$ direction as represented by the white arrows, and a magnetic field $H$ is applied along the $z$ direction. $L$ is the distance between the injector and detector electrodes. The spin injection point is shown as a black solid dot at the infinitesimal interval $[y_0, y_0 + dy_0]$. (a) $S(r)$ represents the spin density at $r$. (b) Mirror image injections are also shown as open dots. $t_{SOI}$ is the thickness of the Si channel fabricated on a SOI substrate.

The magnetizations of both electrodes point to the $x$ direction, and the spin polarizations of the injector and detector are $P^I$ and $P^D$, respectively. Here we consider that the electron spins are injected from the infinitesimal interval $[y_0, y_0 + dy_0]$ at $(x, y, z) = (x, y_0, 0)$. In this situation, the system in the channel has the cylindrical symmetry whose center lies on the line $y = y_0$. Thus, the impulse response function in 2-D $f^{2D}(r, H)$ is expressed as the function of the distance from the center $r = \sqrt{(y - y_0)^2 + z^2}$. and is calculated by integrating the 3-D impulse response Eq. (S9) along the line $y = y_0$ from $x_0 = -\infty$ to $x_0 = +\infty$.

$$f^{2D}(r,H)dy_0 = \int_{x_0=-\infty}^{\infty} f^{3D}(\sqrt{(x-x_0)^2+r^2})dy_0 dx_0$$

$$= \frac{dy_0}{\pi\lambda_S} \int_0^{\infty} \frac{1}{\sqrt{x^2+r^2}} \exp\left(-\alpha\sqrt{x^2+r^2}\right)dx$$

$$= \frac{dy_0}{\pi\lambda_S} K_0(\alpha r), \tag{S11}$$

where $K_0$ is the modified Bessel's function of the second kind. Using Eqs. (S8) and (S11), the spin accumulation signal observed at a distance $r$ is given by

$$\Delta V(r,H) = \Delta V_0^{spin} \text{Re}\left[f^{2D}(r,H)dy_0\right]. \tag{S12}$$

### S3. Derivation of the 2-D impulse response for the lateral device

In this section, we derive the impulse response for a thin body SOI channel with thickness $t_{SOI}$, as shown in Fig. S2(b). Boundary conditions are the spin injection continuity (Eq. (S5)) and no-flux across the planes $z=0$ and $z=t_{SOI}$;

$$\left[-D\frac{d}{dz}\hat{S}(r)\right]_{z=0} = \left[-D\frac{d}{dz}\hat{S}(r)\right]_{z=t_{SOI}} = 0, \tag{S13}$$

These conditions are satisfied by locating mirror images of the spin injection at $z=2t_{SOI}n$ (where $n = 0, \pm 1, \pm 2...$) which are shown as open dots in Fig. S2(b). The spin density at the point $(x, y, 0)$ is given by the discrete infinite series of $\hat{S}(r)$ using $r=\sqrt{(y-y_0)^2+(2t_{SOI}n)^2}$. Thus, the impulse response function $f^{SOI}$ for the spin transport in a SOI channel is given by following discrete infinite series of Eq. (S11);

$$f^{SOI}(y-y_0,H)dy_0 = \sum_{n=-\infty}^{\infty} f^{2D}(\sqrt{(y-y_0)^2+(2t_{SOI}n)^2},H)\,dy_0$$

$$= \frac{dy_0}{\pi\lambda_S}\sum_{n=-\infty}^{\infty} K_0\left(\alpha\sqrt{(y-y_0)^2+(2t_{SOI}n)^2}\right). \tag{S14}$$

When $|2\alpha t_{SOI}| \ll 1$ or $t_{SOI} \ll \lambda_S$, the discrete infinite series can be approximated by the continuous integration. Thus,

$$f^{SOI}(y-y_0,H)dy_0 \approx \frac{dy_0}{\pi\lambda_S}\int_{-\infty}^{\infty}K_0(\alpha\sqrt{(y-y_0)^2+z^2})\frac{dz}{2t_{SOI}}$$

$$=\frac{1}{2}\frac{dy_0}{t_{SOI}}\frac{1}{\sqrt{1+i\gamma H\tau_S}}\exp(-\alpha|y-y_0|) \qquad (S15)$$

Here, we used the integral formula of Bessel's function $\int_0^\infty K_0(\sqrt{a^2+x^2})dx = \frac{\pi}{2}\exp(-a)$ (Re$[a]>0$). The factor $1/2$ in Eq. (S15) comes from the fact that the injected electron spins diffuse to both the left side and right side in the SOI channel along the $y$ direction. Finally, using Eq. (S8), the potential of the detector electrode at the point $(x,y,0)$ in the lateral device is given by

$$\Delta V(y-y_0,H) = \Delta V_0^{spin}\text{Re}\left[f^{SOI}(y-y_0,H)dy_0\right]. \qquad (S16)$$

## S4. Three-terminal Hanle signals in a vertical device structure

The three terminal Hanle signal (N-3TH) $\Delta V^{\text{N-3TH(vertical)}}$ in the vertical device is given by the multiple integration of Eq. (S12) over the injector electrode length [S3];

$$\Delta V^{\text{N-3TH(vertical)}}(H) = \Delta V_0^{spin}\text{Re}\left[\int_0^{l^I}\frac{dy}{l^I}\int_0^{l^I}f^{2D}(|y-y_0|,H)dy_0\right]. \qquad (S17)$$

The first integration is the averaging of the spin detection voltage over the injector electrode, and the second integration is the convolution of the spin injections over the injector electrode. When $l^I \gg \lambda_S$, the range of the second integration in Eq. (S17) can be approximately replaced by $(-\infty,+\infty)$. In that case, the detection voltage is identical for any $y$, in other words, it is not needed to be averaged, so the first averaging integration term is omitted.

$$\Delta V^{3\text{TH(vertical)}}(H) \approx \Delta V_0^{spin} \text{Re}\left[\int_{-\infty}^{\infty} f^{2D}(|y_0|, H) dy_0\right]$$

$$= \Delta V_0^{spin} \text{Re}\left[\frac{2}{\pi \lambda_S} \int_0^{\infty} K_0(\alpha y_0) dy_0\right]$$

$$= \Delta V_0^{spin} \text{Re}\left[\frac{1}{\sqrt{1 + i\gamma H \tau_S}}\right]$$

$$= \Delta V_0^{spin} \sqrt{\frac{1 + \sqrt{1 + (\gamma H \tau_S)^2}}{2 + 2(\gamma H \tau_S)^2}}. \tag{S18}$$

It is notable that this formula is the same as the result derived from the 1-D spin diffusion model, and twice as large as the conventional function [S5]. When the magnetic field angle $\theta$ is varied between the $x$ and $z$ directions, namely, $\mathbf{H} = (H\cos\theta, 0, H\sin\theta)$, Eq. (S18) is modified to

$$\Delta V^{N-3TH(vertical)}(H, \theta) \approx \Delta V_0^{spin} \left[\sqrt{\frac{1 + \sqrt{1 + (\gamma \tau_S H)^2}}{2 + 2(\gamma \tau_S H)^2}} \sin^2\theta + \cos^2\theta\right]. \tag{S19}$$

This is Eq. (2) in the main manuscript.

**S5. Three-terminal and four-terminal Hanle signals in the lateral device**

The three-terminal and four-terminal Hanle signals in the lateral device ($\Delta V^{\text{N-3TH(lateral)}}(H)$ and $\Delta V^{\text{4TH(lateral)}}(H)$) are given by the multiple integration [S3,S4] of Eq. (S16) over the lengths of the injector and detector electrodes;

$$\Delta V^{4\text{TH(lateral)}}(H) = \Delta V_0^{spin} \text{Re}\left[\int_{l^I+L}^{l^I+L+l^D} \frac{dy}{l^D} \int_0^{l^I} f^{\text{SOI}}(|y-y_0|, H) dy_0\right]$$

$$= \frac{1}{2} \Delta V_0^{spin} \frac{\lambda_S}{t_{\text{SOI}}} \text{Re}\left[\frac{1}{1+i\gamma H \tau_S} \exp(-\alpha L) \frac{1}{\alpha l^D}(1-\exp(-\alpha l^D))(1-\exp(-\alpha l^I))\right],$$

$$\tag{S20}$$

$$\Delta V^{\text{N-3TH(lateral)}}(H) = \Delta V_0^{spin} \text{Re}\left[\int_0^{l^I}\frac{dy}{l^I}\int_0^{l^I} f^{\text{SOI}}(|y-y_0|,H)\,dy_0\right]$$

$$= \Delta V_0^{spin}\frac{\lambda_S}{t_{\text{SOI}}}\text{Re}\left[\frac{1}{1+i\gamma H\tau_S}\left\{1-\frac{1-\exp(-\alpha l^I)}{\alpha l^I}\right\}\right]. \tag{S21}$$

Eqs. (S20) and (S21) are Eqs. (4) and (5) in the main manuscript, respectively.

Below, we show the long- and short-electrode limits of Eqs. (S20) and (S21), and compare them with the functions in previous reports by other groups. When both the lengths of the injector and detector electrodes are shorter than the spin diffusion length, i.e., $l^D \ll \lambda_S$ and $l^I \ll \lambda_S$, Eqs. (S20) and (S21) can be approximated as follows;

$$\Delta V^{\text{4TH(lateral)}}(H) \approx \frac{1}{2}\Delta V_0^{spin}\frac{l^I}{t_{\text{SOI}}}\text{Re}\left[\frac{1}{\sqrt{1+i\gamma H\tau_S}}\exp(-\alpha L)\right]$$

$$= \frac{1}{2}\Delta V_0^{spin}\frac{l^I}{t_{\text{SOI}}}\left(1+(\gamma H\tau_S)^2\right)^{-1/4}\exp\left[-\frac{L}{\lambda_S}\sqrt{\frac{\sqrt{1+(\gamma H\tau_S)^2}+1}{2}}\right]$$

$$\times \cos\left[\frac{\tan^{-1}(\gamma H\tau_S)}{2}+\frac{L}{\lambda_S}\sqrt{\frac{\sqrt{1+(\gamma H\tau_S)^2}-1}{2}}\right], \tag{S22}$$

$$\Delta V^{\text{N-3TH(lateral)}}(H) \approx \frac{1}{2}\Delta V_0^{spin}\frac{l^I}{t_{\text{SOI}}}\sqrt{\frac{1+\sqrt{1+(\gamma H\tau_S)^2}}{2+2(\gamma H\tau_S)^2}} \quad (l^I, l^D \ll \lambda_S). \tag{S23}$$

The term $l^I/t_{\text{SOI}}$ was previously introduced as the geometrical factor in refs. [S2] and [S9]. Note that Eqs. (S22) and (S23) are independent of $l^I$ because the current density $J$ in $\Delta V_0^{spin}$ contains $1/l^I$. On the other hand, when $l^D, l^I \gg \lambda_S$, Eqs. (S20) and (S21) can be approximately written as follows;

$$\Delta V^{\text{4TH(lateral)}}(H) \approx \frac{1}{2}\Delta V_0^{spin}\frac{\lambda_S}{t_{\text{SOI}}}\text{Re}\left[\frac{1}{1+i\gamma H\tau_S}\frac{1}{\alpha l^D}\exp(-\alpha L)\right], \tag{S24}$$

$$\Delta V^{\text{3TH(lateral)}}(H) \approx \Delta V_0^{spin}\frac{\lambda_S}{t_{\text{SOI}}}\frac{1}{1+(\gamma H\tau_S)^2}. \tag{S25}$$

The term $\lambda/t_{SOI}$ was previously introduced as the geometrical factor in ref. [S6]. In Fig. S3(a), the calculation results of Eq. (S20) using $l^I/\lambda_S = 5$ and of Eq. (S22) are plotted as a solid curve and a broken curve, respectively.

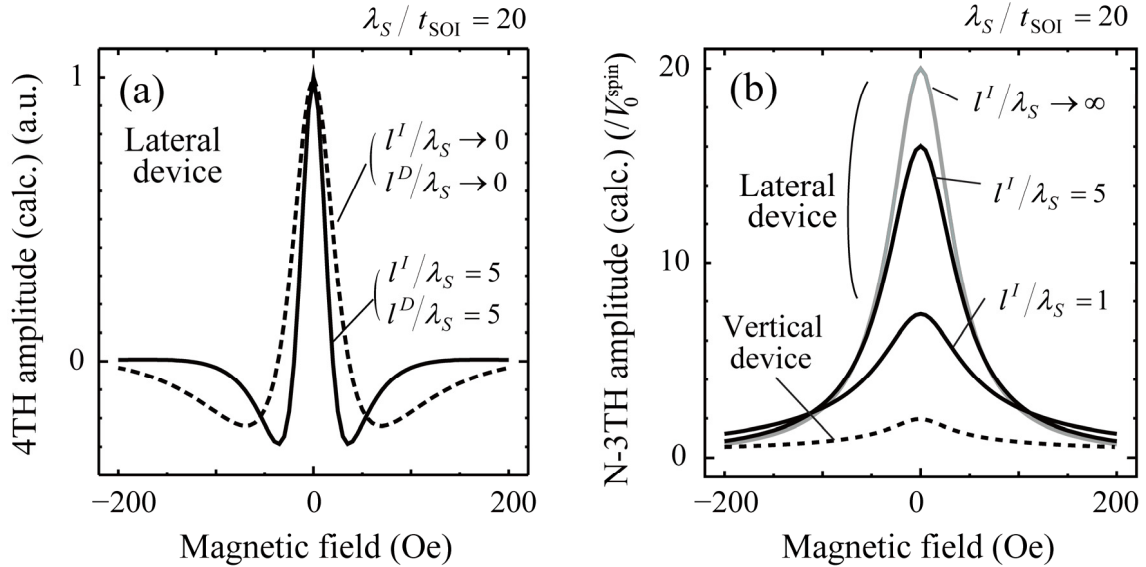

Figure S3  (a) Calculated line shape of the 4TH signals using Eq. (S20) with $l^D/\lambda_S = l^I/\lambda_S = 5$ (solid line) and $l^D/\lambda_S = l^I/\lambda_S \to 0$ (dashed line, Eq. (S22)). (b) Solid curves are calculated N-3TH signals in lateral devices with $\lambda_S/t_{SOI} = 20$ using Eqs. (S21) and various $l^I/\lambda_S$; $l^I/\lambda_S \to \infty$, $l^I/\lambda_S = 5$, and $l^I/\lambda_S = 1$. Broken curve is calculated line shape of the N-3TH signal in a vertical device using Eq. (S18).

As seen in Eq. (S20), the amplitude of $\Delta V^{4TH(lateral)}$ decreases by the factor $1/\alpha l^D$ as the detector electrode length increases. Also, the linewidth becomes narrower as both the injector and detector electrode lengths increase. Here, this effect is called electrode averaging effect (EAE), which is more pronounced when $l^I, l^D \gg \lambda_S$. On the other hand, in Fig. S3(b), the calculation results of Eq. (S21) are plotted as solid curves using $\lambda_S/t_{SOI} = 20$ and various injector electrode lengths; $l^I/\lambda_S \to \infty$ (Eq. (S25)), $l^I/\lambda_S = 5$, and $l^I/\lambda_S = 1$. Eq. (S18) is also shown in Fig. S3(b) as a broken curve. Comparing the N-3TH signal in the vertical (broken line) and the lateral device (gray line) with $l^I, l^D \gg \lambda_S$, the amplitude of the signal in the lateral device is larger than that in the vertical device by the factor $\lambda_S/t_{SOI}$. This is because the diffusion of the injected electron spins towards the backside of the channel is prohibited by the buried oxide (BOX) layer and thus the spins are confined in the thin channel (thickness $t_{SOI}$) in the lateral device. This is called channel confinement effect (CCE), and it is represented by the term $\lambda_S/t_{SOI}$ in Eqs. (S24) and (S25) when $l^D \gg \lambda_S$, and $l^I/t_{SOI}$ in Eqs. (S22) and (S23) when $l^I \ll \lambda_S$. Considering the N-3TH signals in the lateral devices, as $l^I/\lambda_S$ decreases, the linewidth and amplitude of the signal becomes broader and lower, respectively (Fig. S3(b)). This is because the diffusion of electron spins to the both left and right sides of the channel becomes more dominant as the injector length becomes shorter, in other words, CCE is weakened.

Comparing Eqs. (S22) and (S23) with the previously reported functions (such as Eq. (3) in ref. [S5]), their magnetic-field dependence is the same as Eq. (S22) and (S23). However, their amplitudes are different because CCE is not taken into account in the functions in ref. [S5]. On the other hand, comparing Eqs. (S24) and (S25) with the previously reported functions which take CCE into account (such as Eq. (1) in ref. [S6] and Eq. (11) in ref. [S7]), their geometrical factors are the same as Eq. (S24) and (S25). However, their magnetic-field dependence in ref. [S6] and [S7] is different because EAE is not taken into account in their functions. Since Eqs. (S17), (S20), and (S21) precisely take into account both CCE and EAE, the estimated spin lifetime and spin polarization in the vertical and lateral devices are almost identical as shown in tables 1 and 2 in the main manuscript.

**S6. Bias dependence of N-3TH signals in degenerated semiconductors**

In the three-terminal measurement, N-3TH signals are observed only in the spin extraction regime, but not in the spin injection regime. The same result was also reported by other groups, however, the origin has not been clearly presented [S8, S9]. To understand this result, the depletion layer in Si formed nearby the Si/MgO interface must be considered. In the spin extraction regime, the applied voltage $V_0$ at the Fe/MgO/Si junction is almost the same as the voltage drop of MgO, $V_{MgO}$ (see Fig. S4(a)). On the contrary, in the spin injection regime, the applied voltage $V_0$ at the Fe/MgO/Si junction is divided into two parts; voltage drop of MgO, $V_{MgO}$, and that of the Si depletion layer, $V_{SC} = \frac{1}{2} E_S L_D$, where $E_S$ is the maximum electric field at the Si surface and $L_D$ is the depletion layer thickness (see Fig. S4(b)). Using $V_0 = V_{MgO} + V_{SC}$, $\varepsilon_{Si} E_S = \varepsilon_{MgO} V_{MgO} / t_{MgO}$, and $L_D = \sqrt{\frac{2\varepsilon_{Si}\varepsilon_0 V_{Si}}{qN_D}}$, where $\varepsilon_{Si}$ = 12 and $\varepsilon_{MgO}$ = 9 are relative dielectric constant of Si and MgO, respectively, $\varepsilon_0$ is the vacuum dielectric constant, $t_{MgO}$ = 0.8 nm, $N_D$ = 8×10$^{19}$ cm$^{-3}$, and $V_0$ = 0.84 V, we estimated $E_S$ = 3.4 MV/cm. For simplicity, we solve the 1-D spin drift-diffusion equation [S8] under the condition that the electric field in the depletion layer is constant at $E_S$ in the half length of the $L_D$ and no field in other regions, that is

$$E(z) = \begin{cases} -E_S & (0 \leq z \leq L_D/2) \\ 0 & (L_D/2 \leq z) \end{cases} \quad \text{(S26)}$$

The drift-diffusion equation is given by

$$D\frac{d^2}{dz^2}\hat{S} - \mu E(z)\frac{d}{dz}\hat{S} - (\frac{1}{\tau_S} + i\omega)\hat{S} + \frac{P^I J}{q}\delta(z) = 0, \quad \text{(S27)}$$

where $\mu$ is the mobility of electrons. Using the following boundary conditions;

$$-D\frac{d}{dz}\hat{S}(z=0) = \frac{P^I J}{q}, \quad \text{(S28a)}$$

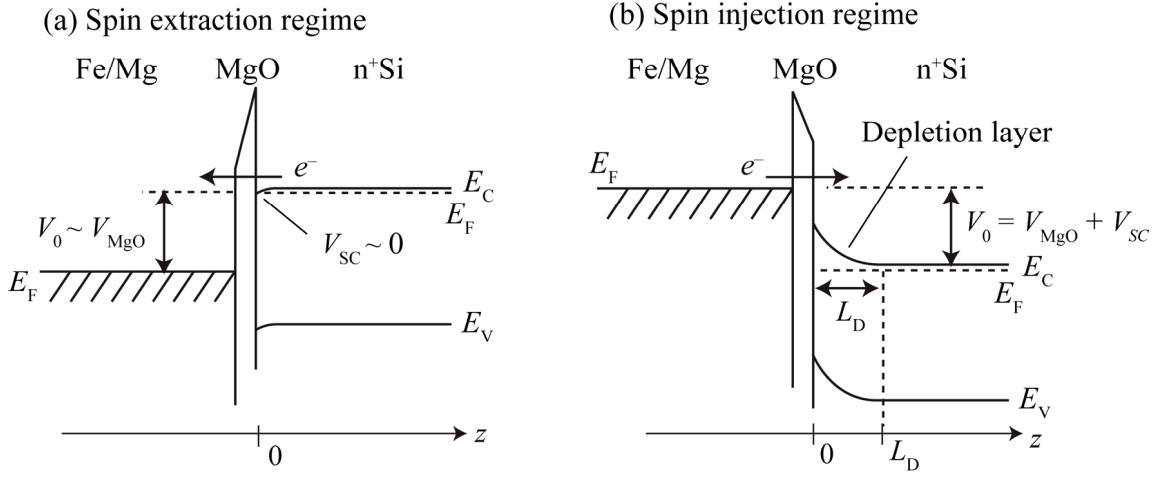

Figure S4  Schematic energy band diagrams of a Fe/Mg/MgO/$n^+$-Si structure in the (a) spin extraction and (b) spin injection regimes.  $E_C, E_F$, and $E_V$ represent the energy of the conduction band bottom, Fermi level, and valence band top in Si, respectively.  $V_{MgO}$, and $V_{SC}$ represent the voltage drops of the MgO layer and the Si depletion layer, respectively.  $V_0 = V_{MgO} + V_{SC}$ is the applied voltage at the Fe/Mg/MgO/Si junction.  The $z$ axis is defined as the depth direction from MgO to Si whose origin is at the MgO/Si interface.  In the spin extraction regime, $V_0 \sim V_{MgO}$ because $V_{SC} \sim 0$.

$$-D\frac{d}{dz}\hat{S}(z = \frac{L_D}{2} - 0) = -D\frac{d}{dz}\hat{S}(z = \frac{L_D}{2} + 0), \tag{S28b}$$

$$-D\frac{d}{dz}\hat{S}(z \to \infty) = 0, \tag{S28c}$$

the solutions of Eq. (S27) are obtained,

$$\hat{S}(z) = \begin{cases} \dfrac{P_I J}{qD}\dfrac{1}{\beta}\exp(-\beta z)u(z) & (0 \le z \le L_D) \\[1em] \dfrac{P_I J}{qD}\dfrac{\exp(-\beta L_D/2)}{\alpha \exp(-\alpha L_D/2)}\exp(-\alpha z)u(z) & (L_D \le z) \end{cases} \tag{S29}$$

, where $\alpha = \dfrac{\sqrt{1+i\gamma H \tau_S}}{\lambda_S}$ and $\beta = \dfrac{\sqrt{(\mu E_S \tau_S/2\lambda_S)^2 + 1 + i\gamma H \tau_S} - \mu E_S \tau_S/2\lambda_S}{\lambda_S}$.

Finally, the N-3TH signal is obtained,

$$\Delta V^{N-3TH}(H, E_S) = V_0^{spin} \operatorname{Re}\left[\left(\sqrt{\left(\frac{\mu E_S \tau_S}{2\lambda_S}\right)^2 + 1 + i\gamma H \tau_S} - \frac{\mu E_S \tau_S}{2\lambda_S}\right)^{-1}\right]. \quad \text{(S30)}$$

In the spin extraction regime, the electric field in Si is negligible as mentioned in the main manuscript. Thus, Eq. (18) is applicable. On the contrary, in the spin injection regime, since the electric field in the depletion layer must be considered, Eq. (S30) must be used. Figures S5(a) and (b) show the amplitude and linewidth of the N-3TH signals, respectively, using Eqs. (S18) and (S30) with $\mu$ = 100 cm$^2$/Vs, $\tau_S$ = 2 ns, and $\lambda_S$ = 1 μm. In Fig. S5(a), our experimental results are also shown by open circles. The amplitude of the N-3TH signals in the spin extraction regime ($V_0 < 0$) linearly increases with $V_0$, whereas the linewidth is constant. On the contrary, in the spin injection regime ($V_0 > 0$), the amplitude of N-3TH signals is much smaller than that in the extraction regime and the linewidth is broadened with increasing $V_0$; this is because the spin-polarized electrons in the depletion layer of the Si channel are drifted away from the interface by the electric field. When the linewidth of the N-3TH signal is broadened and becomes comparable to that of the B-3TH signal, the N-3TH signal cannot be distinguished from the B-3TH signal. These results are consistent with our experimental results as mentioned in the main manuscript, as well as the previously reported results [S8, S9]. Therefore, the depletion layer must be considered even in degenerated semiconductors.

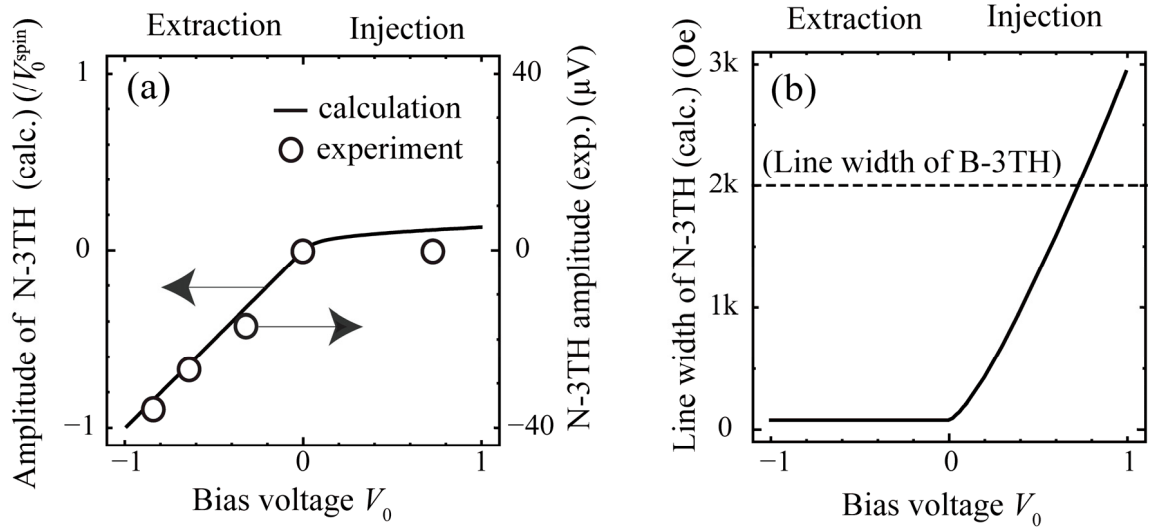

Figure S5  Calculated (a) amplitude and (b) linewidth of the N-3TH signal using Eq. (S18) for the negative bias (the spin extraction regime), and Eq. (S30) for the positive bais (the spin injection regime).   Experimental results of the vertical device with $t_{MgO}$ = 0.8 nm and $t_{Mg}$ = 1.0 nm are also plotted by the open circles in Fig. S5(a).   In Fig. S5(b), the linewidth of the B-3TH signals (~ 2 kOe) is shown as a broken line.